\documentclass{ws-procs9x6}
 \def\@cite#1{\textsuperscript{[#1]}}

\def\be{\begin{equation}}
\def\ee{\end{equation}}

\newcommand{\bea}{\begin{eqnarray}}
\newcommand{\eea}{\end{eqnarray}}
\newcommand{\bwt}{\begin{widetext}}
\newcommand{\ewt}{\end{widetext}}

\def\eed{\end{document}}

\def\al{\alpha}

\def\m_z{m_{\textrm {Z}}}
\def\al{\alpha}
\def\be{\beta}

\def\al{{\alpha}}

\begin{document}

\title{Constrained Electroweak Chiral Lagrangian}

\author{Qi-Shu Yan}

\address{NCTS (Theory Division),
101, Section 2 Kuang Fu Road, Hsinchu, Taiwan\\
$^*$E-mail: yanqs@phys.nthu.edu.tw
}

\begin{abstract}
We update the uncertainty analysis on $S$ parameter of
the electroweak chiral Lagrangian (EWCL) by including the LEP-II W pair production
data. We find that experimental data still allow
a positive $S^{EXP}(1\textrm{TeV})$.
\end{abstract}

%\keywords{Electroweak chiral Lagrangian, triple gauge couplings}

\bodymatter

\section{Introduction}
Technicolor models are one of the natural candidates 
beyond the standard model \cite{hill}.
But it is widely said that the Technicolor models are
ruled out by electroweak precision data. The
claim \cite{pandt} is established in two logical steps: 1) by extrapolating the
electroweak precision data from $\mu=m_Z$ to 1 TeV (1 TeV is argued as
the scale of compositeness) by perturbation calculation with only $6$ quadratic operators
of the EWCL, it was found that the value of
the parameter $S$ is negative. In our fit, 
when triple gauge coupling (TGC) effects are not included, $S$ is determined as
\bea
S^{EXP}(1 \textrm{TeV}) = -0.17 \pm 0.10\,,
\eea
 2) by using the unsubtracted dispersion relations with the assumption
 of custodial symmetry and vector meson dominance
and the low energy hadronic QCD data of $\rho$ and $a_1$ mesons as
input, and by scaling up the value of $S$ to 1 TeV, it was found that the
value of $S$ parameter is positive \cite{pandt}, which is given as
\bea
S^{TH}(1 \textrm{TeV})=0.3 \frac{N_{TF} N_{TC}}{6}\,.
\eea
The discrepancy between $S^{EXP}$ and $S^{TH}$
is at least $3 \sigma$, which 
is interpreted as an evidence that Technicolor
models are ruled out by precision data.

In this article we summarize our study on
the uncertainty analysis in $S^{EXP}(1 \textrm{TeV})$ \cite{full}
by extending the $6$-operator analysis to the $14$-operator one. 
Part of results was reported in \cite{our}. Our result show that the uncertainty
induced by TGC measurement dominates the error bar of $S^{EXP}(1\textrm{TeV})$
and experimental data still allows
positive $S^{EXP}(1\textrm{TeV})$ in the parameter space.
Therefore we argue that it is premature to claim that the electroweak 
precision data has ruled out Technicolor models.

\section{Our knowledge on the chiral coefficients}
We follow the standard analysis of chiral 
Lagrangian method \cite{Gasser:1983yg,Harada:2003jx} and 
include $14$ operators up to mass dimension 
four in the EWCL \cite{Appelquist:1993ka}. 
Our study extends the RGE analysis of 
Bagger {\it et. al.} \cite{Bagger},
who have considered the effects of  
$6$ out of the $14$ operators.

The $14$ gauge invariant operators constructed in the EWCL
are supposed to describe EWSB models defined at $1 \textrm{TeV}$ 
in a model independent way, either strong or weak interaction models.
Below we describe how to determine $14$ chiral coefficients of 
the EWCL in our analysis at $\mu=m_Z$.
Three of six two-point chiral coefficients $g$, $g'$, and $v$, 
are determined by the following inputs 
$1/\al_{\textrm{em}}(\m_z)= 128.74$, $\m_z=91.18$ GeV, and
$G_F=1.16637 \times 10^{-5}$ GeV$^{-2}$. They are assumed
to be free from error bars.
The next three two-point-function chiral coefficients $\al_1$,
$\al_0$, and $\al_8$ are extracted from 
Z-pole data,
precise W mass measurement, 
and top quark mass measurement (Here we use data of PDG2004).
We perform the analysis with three best measured 
quantities $m_\textrm{W}= 80.425\pm 0.038$ GeV, 
 $\sin^2\theta_W^{\textrm{eff}}=0.23147\pm 0.00017$ 
and the leptonic decay width of $Z$, $\Gamma_\ell =83.984\pm 0.086$ MeV
for the $S,\, T$ and $U$ fit. We take 
$m_{\textrm{t}}=175$ GeV in our $S$, $T$, and $U$ fit.
 
The central values with  $1\sigma$ errors of the $S$, $T$, and $U$ parameters 
are found as
\begin{equation}
\begin{array}{rl}
S(m_Z) = &(-0.06\pm 0.11)\\
T(m_Z) = &(-0.08\pm 0.14)\\
U(m_Z) = &(+0.17\pm 0.15)
\end{array}
\rho_{co.} =\left(\begin{matrix}
1\,\,\, & & \cr 
.9 & 1\,\,\, & \cr 
-.4&-.6& 1\cr
\end{matrix}
\right)
\label{stfit}
\end{equation}
which roughly agrees with \cite{Bagger}.

The relations among two-point chiral coefficients $\al_1$, $\alpha_0$, 
and $\al_8$ of EWCL with the $STU$ parameters are found to be
\begin{eqnarray}
&&\hskip -0.5 cm 
\begin{array}{l}
\al_1 (\mu)= - \frac{1}{16 \pi}\,  S(\mu)  , \\
\al_0 (\mu)=   \frac{1}{2} \, \al(\mu)_{\textrm{EM}}^{} \, T(\mu) , \\
\al_8 (\mu)= - \frac{1}{16 \pi} \, U(\mu)\,.
\label{stu}
\end{array}
\end{eqnarray}

From Eqs. (\ref{stfit}-\ref{stu}),
$\al_1(m_Z)$, $\al_0(m_Z)$, and $\al_8(m_Z)$ are determined as
\begin{equation}
\begin{array}{rl}
\al_1(m_Z) = &(+0.13\pm 0.21) \times 10^{-2}\\
\al_0(m_Z) = &(-0.03\pm 0.05) \times 10^{-2}\\
\al_8(m_Z) = &(-0.35\pm 0.29) \times 10^{-2}
\end{array}
\label{eq:al108}
\end{equation}

Three three-point chiral coefficients $\al_2$, $\al_3$ 
and $\al_9$ are extracted from the LEP-II
measurements via the process $e^+ e^- \rightarrow W^+ W^-$ \cite{TGCsum}.
The experimental observables 
of anomalous TGC \cite{Hagiwara:1986vm} between 
$\delta k_\gamma$, $\delta k_Z$, $\delta g_1^Z$,  
and three-point chiral coefficients,  $\al_2$, $\al_3$, $\al_9$ ,
can be simplified as 
\begin{eqnarray}
&&\hskip -0.5 cm 
\begin{array}{l}
\delta k_\gamma =  (\al_2 + \al_3 + \al_9) g^2 \,, \\
\delta k_Z =  (\al_3 + \al_9) g^2 - \al_2 {g^{\prime}}^2\,,  \\
\delta g_1^Z =   \al_3 (g^2+{g^\prime}^2 )\,\,.
\end{array}
\end{eqnarray}

There is no experimental data relaxing the custodial symmetry 
 except L3 collaboration
\cite{L3group}  from where we take $\delta  k_Z = -0.076\pm 0.064$ 
as one of the inputs. Other inputs $\delta  k_\gamma =-0.027 \pm 0.045 $ and 
$\delta  g^Z_1 = -0.016 \pm 0.022$ are taken from LEP Electroweak working 
group \cite{LEPEWWG, Heister:2001qt}. 
We found TGC errors are quite large as reported in D0 
collaboration \cite{Abazov:2005ys} at Tevatron.
Because of this fact, we use LEP data in our analysis.

We also relax the custodial $SU(2)$ gauge 
symmetry as it is natural in the framework
of the EWCL to have a non-vanishing $\al_9$ if the
underlying dynamics break this symmetry explicitly \cite{SSVZ}.  
By assuming these data are extracted independently,  
we can obtain three-point chiral coefficients
as 
\begin{equation}
\label{3pfit}
\begin{array}{rl}
\al_2(m_Z) =  \!\!\!&(+0.09\pm 0.14)\\
\al_3(m_Z) =  \!\!\!&(-0.03\pm 0.04)\\
\al_9(m_Z) =  \!\!\!&(-0.12\pm 0.12)
\end{array}\!
\rho_{co.}\! \! = \! \! \left(\begin{matrix}
1 \,\,\,& & \cr 
0 & 1 \,\,\,& \cr 
-.7&-.3&1\,\,\,\cr
\end{matrix}\!\!
\right)\!\! .
\end{equation}
We observe that $\al_3(m_Z)$ is more tightly constrained 
than $\al_2(m_Z)$ and $\al_9(m_Z)$.
In our numerical analysis, we consider the two-parameter fit
data from L3 collaboration and this combined data as two scenarios
to show the effects of TGC to $S^{EXP}$.

There is no experimental data to bound
5 four-point chiral coefficients, which 
usually are assumed to be of order one.
Partial wave unitary bounds of longitudinal vector boson
scattering processes can be used to put bounds on the magnitude
of those chiral coefficients. 
We use the following five $J=0$ channels to bound $5$ chiral 
coefficients ($\Lambda$ is arbitrary, which should correspond to
the UV cutoff of the EWCL),
$\al_4$, $\al_5$, $\al_6$, $\al_7$, and $\al_{10}$:
\begin{eqnarray}
&&\hskip -0.5 cm 
\begin{array}{rl}
|4 \al_4 + 2\al_5 |  <&  {3 \pi} \frac{v^4 }{ \Lambda^4},\\
| 3 \al_4 + 4 \al_5 |  <& {3 \pi } \frac{v^4 }{ \Lambda^4},\\
|\al_4 + \al_6 + 3 (\al_5 + \al_7) | < & {3 \pi } \frac{v^4 }{ \Lambda^4},\\
|2 (\al_4 +\al_6) + \al_5  +\al_7 |  < & {3 \pi } \frac{v^4 }{ \Lambda^4},\\
| \al_4 + \al_5 + 2 (\al_6 + \al_7 + \al_{10}) |  < & 
\frac{6 \pi }{5 } \frac{v^4 }{ \Lambda^4},
\end{array}
\label{quartic-bounds}
\end{eqnarray}
%\end{array}
%\end{equation}
where bounds are obtained from $W^+_L W^+_L \rightarrow W^+_L W^+_L$,
$W^+_L W^-_L \rightarrow W^+_L W^-_L$, $W^+_L W^-_L \rightarrow Z_L Z_L$,
$W^+_L Z_L   \rightarrow W^+_L Z_L $, and $Z_L Z_L \rightarrow Z_L Z_L$, 
respectively. Quartic gauge couplings will not contribute to $S$ parameter,
but do affect $T$ parameter. For completeness, we list these theorectical bounds
here.

\section{Uncertainty in $S^{EXP}(\Lambda)$ parameter}
In the framework of effective field theory method \cite{georgi}, 
all chiral coefficients
of the EWCL depend on renormalization scale $\mu$.
$S(\Lambda)$, $T(\Lambda)$, and $U(\Lambda)$ are  
values of parameters $S$, $T$, 
and $U$ at the matching scale $\Lambda$,
where the EWCL matches with 
fundamental theories, Technicolor models, 
extra dimension models, Higgsless models, etc.
In the perturbation method, theoretically, 
the $S(m_Z)$, $T(m_Z)$, and $U(m_Z)$ can be connected
with the $S(\Lambda)$, $T(\Lambda)$, and $U(\Lambda)$
by improved renormalization group equations \cite{full}.
 
The effect of TGC measurement is depicted 
on $S(\Lambda)-T(\Lambda)$  plane as shown 
in Fig. 1(a) and Fig. 1(b).
We highlight some features of these two figures.
{\bf (1)}  In Fig. 1(a), in absence of TGC contribution (dashed-line contours),  $S(\Lambda)$  
becomes more negative as $\Lambda$ increases with reference to the reference LEP-I fit 
contour at $\Lambda = m_Z$. This is roughly in agreement with
the observation of Ref. \cite{Bagger}, Ref. \cite{Peskin:2001rw}, and PDG \cite{pdg2004}. 
Two-parameter fit data with custodial 
symmetry condition from L3 collaboration shows that
$S(\Lambda)$ is driven to the positive value region (the solid line).
The central value of $S^{EXP}(1 \textrm{TeV})$ is $0.8$. Furthermore, the error
bars of $S-T$ are greatly enlarged by the uncertainty of TGC measurements.

{\bf (2)} In Fig. 1(b), without the custodial symmetry assumption on the
experimental data, as given in Eq. (\ref{3pfit}),
inclusion of TGC contribution makes the central value of $S(\Lambda)$ almost unchanged. 
Meanwhile, the error bars
of $S-T$ are enlarged at least by a factor of 2.

\begin{figure}[t]
\centerline{
\epsfxsize= 6.0 cm\epsfysize=5.5cm
 \epsfbox{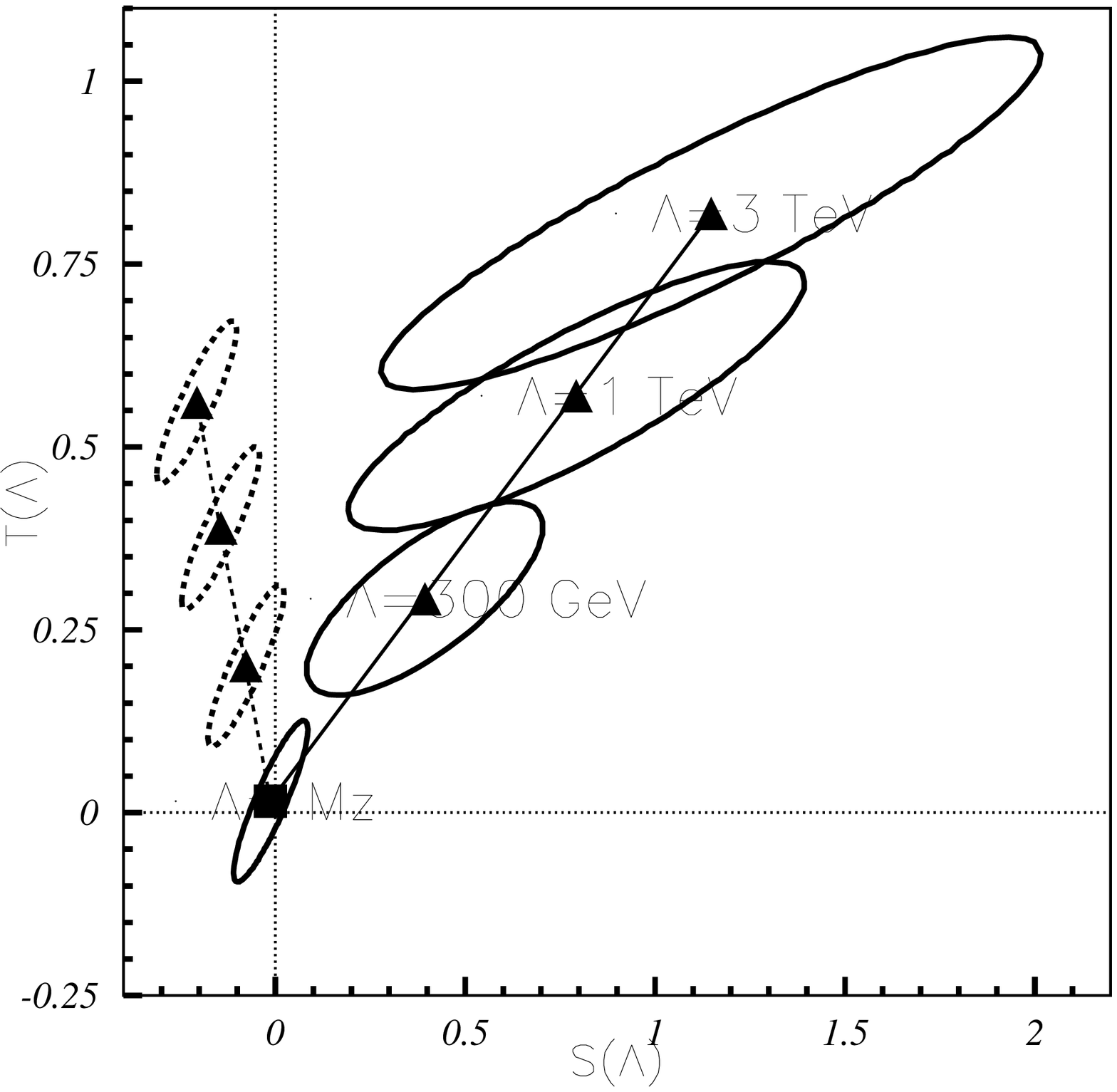}
        \hspace*{0.1cm}
\epsfxsize=6.0 cm\epsfysize=5.5cm
                     \epsfbox{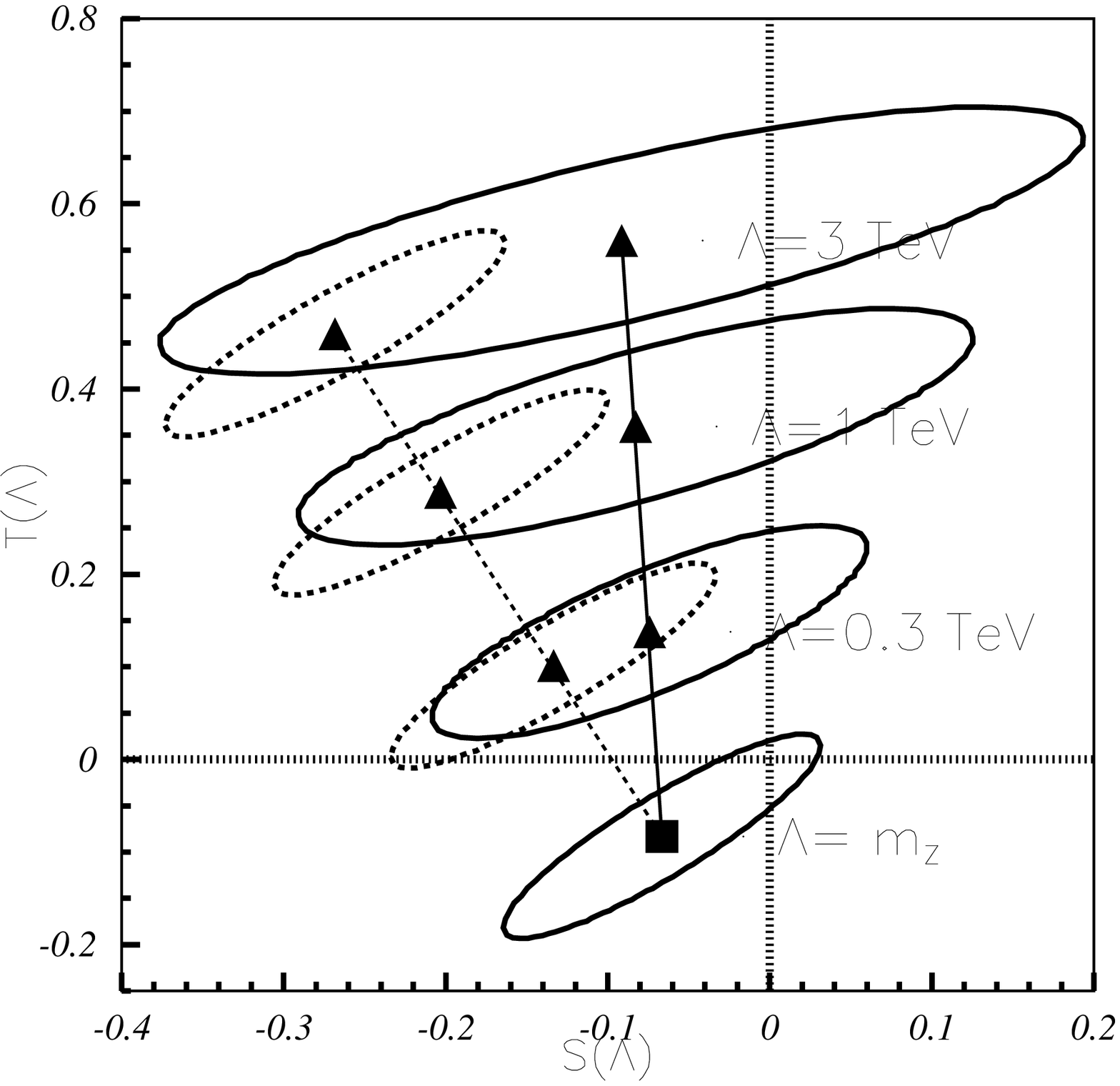}}
\vskip -.25cm
\hskip 2.2 cm {\bf ( a) } \hskip 5.8cm {\bf (b)}
\caption{\it
$S(\Lambda)-T(\Lambda)$ contours at $\Lambda=\m_z$, $300$ GeV,
$1$ TeV, and $3$ TeV, respectively. TGC uncertainty is included in solid line 
contours  while not included in dashed line. Fig1(a) corresponds to
the L3 two-parameter fit data. The parameter $U(m_Z)$ is taken 
as its best fit value, $U(m_Z)=+0.17$. Fig1(b) corresponds to the
combined data. The parameter $U(m_Z)$ is taken
as its best fit value, $U(m_Z)=0.00$. 
}
\label{fig2}
\end{figure}

Fig. 1 clearly demonstrate a fact
that the TGC measurement affect the value of $S^{EXP}(\Lambda)$ significantly.

\section{Discussions and Conclusions}
In the unsubtracted
dispersion relation \cite{Gasser:1983yg,pandt},
the fact that $S^{EXP}$ is a running parameter might not be transparent.
In the RGE analysis, 
the quantum fluctuations of active degree of freedoms $S$ parameter to run:
\bea
S^{EXP}(\Lambda) =& S^{EXP}(m_Z) + \beta_{S} \ln \frac{\Lambda}{m_Z}\,.
\label{rgef}
\eea
Active degree of freedoms and new resonances can contribute to $\beta_{S}$ function and affect
the value of $S^{EXP}(\Lambda)$.
We propose a naive subtracted dispersion relation without assuming custodial symmetry
and vector meson dominance, which can read as
\bea
S(q^2) &=& S(q^2=0) - \frac{q^2}{3 \pi} \int_{s > m_z^2}^{\infty} ds \frac{R_{3Y}(s)}{s (s -q^2)}\,.
\label{drf}
\eea
Where $R_{3Y}=-12 \pi Im \Pi^{\prime}_{3Y}$, which is to count the degree of freedoms
belonging to the representations of both $U_Y(1)$ and $SU_L(2)$.
The $S(q^2=0)$ corresponds to the value determined from LEP-I Z-pole data. 
The second term
include the contributions of active degree of freedoms and new resonances, 
either fermionic or bosonic. Wth Eqs. (\ref{rgef}-\ref{drf}),
the equivalence between the description of RGE
and dispersion relation becomes obvious \footnote {We thank Han-Qing Zheng and Kenneth Lane
for comments on this point.}.

Our most conservative numerical result from Fig. 1(b)
can be put as
\begin{equation}
\begin{array}{rl}
S^{EXP}(1 \textrm{TeV}) = &-0.08\pm 0.20\,.
\end{array}
\label{st1tev}
\end{equation}
To obtain these numerical values, we have assumed the
perturbation method is valid from the energy scale $m_Z$ to $1 \textrm{TeV}$. 
When data in PDG2006 is used, the central valus of $S^{EXP}(1\textrm{TeV})$ 
shifts to $-0.02$, which is in agreement with the observation in \cite{Dietrich:2005wk}.
If there exist new
resonances below or near $1 \textrm{TeV}$, the perturbation method
might be invalid before $\Lambda=1 \textrm{TeV}$ and effects of threshold and tail of new resonances
might modified the value of $S(1 \textrm{TeV})$ drastically.
This might occur, 
for example, in the low energy Technicolor model \cite{Lane:2002sm}.
Hence, we stress that both the center values and error bars of
$S(1 \textrm{TeV})$ can only be interpreted as reference
values obtained in perturbation method.

Whether $S$ and $T$ should run in a logarithmic or power way or whether the decoupling
theorem should hold in the process of extrapolating the data from $m_Z$ scale to the
matching scale is still a debatable issue. 
In the analysis of the minimal
standard model with a Higgs \cite{pdg2004}, the model is renormalizable
and Higgs boson
plays the role of regulator. Therefore only logarithmic
terms are taken in the standard global fit.
In the EWCL (a non-renormalizable theory),
when operators beyond the $O(p^2)$ order are included, 
terms proportional to the power
of $\Lambda_{UV}/v$ enter into the radiative corrections.
As the most conservative calculation, we adopted the 
logarithmic running. However, if 
power running is used, 
error bars of $S$ parameters would be much larger than those shown Fig. 1.

One-loop contribution of TGC in our analysis can be attributed as part of
$O(p^6)$ order effects in the standard chiral derivative power counting rule.
There are analysis by including $O(p^6)$ operators in order to 
accommodate data of $e^+ e^- \rightarrow f {\bar f}$
above Z pole, as done in \cite{Barbieri:2004qk}. Even when
these tree level effects of $O(p^6)$ operators are included, near $TeV$
region, the sign of $S^{EXP}(1\textrm{TeV})$ can not change from negative to positive.
Another remarkable fact is that when more operators are introduced
the error bar of $S^{EXP}(1\textrm{TeV})$ becomes a few larger.
But effects of these operators are smaller than those of TGC operators.
 
One may worry about the two-loop contributions of $O(p^2)$, 
which are also part of $O(p^6)$ order effects. However,
due to the two loop suppression
factor, they must be tiny.
Therefore, uncertainty induced by the TGC dominates the error
bar of $S^{EXP}(1\textrm{TeV})$. Our results show that the 
sign of $S(1\textrm{TeV})$ can be changed from negative
to positive by TGC.

It is an open question to construct a realistic model which can
have a large deviation from
the prediction of the SM in TGC sector. 
In the Higgsless model with ideally delocalized fermions 
and the gauge-Higgs unification model in the warped space-time,
it was found that the deviation is small \cite{tgcth}.
The Higgsless model in warped 5D space-time might provide a solution.
\footnote {We thank Kinya Oda mention this to us.}

We show here that electroweak precision data have constrained 
both the oblique parameters STU
significantly and the anomalous TGC considerably.
But, current precision of electroweak data is not 
sufficient to rule out Technicolor models, due to 
the large uncertainty in $\al_2$ and $\al_9$.
Technicolor models can provide
dark matter candidates \cite{Gudnason:2006yj}.
Therefore, in our opinion, Technicolor models are still quite competitive and 
promising as a candidate of
EWSB \cite{Dietrich:2006cm}.

\vskip 0.2cm
\noindent {\bf ACKNOWLEDGMENTS}\\
This work is partially supported by the JSPS fellowship program 
and by NCTS (Hsinchu, Taiwan). 
We would like to thank Ulrich Parzefall for communication on the TGC
measurements, Masaharu Tanabashi and Masayasu Harada 
for stimulating discussions.

\end{document}